\documentclass[final,5p,twocolumn]{elsarticle}

\usepackage{graphicx}
\usepackage{float}
\usepackage{tabularx}
\usepackage{adjustbox}
\usepackage{lipsum}
\usepackage{bbm}
\usepackage[dvipsnames]{xcolor,colortbl}
\usepackage[figurename=Figure, labelsep=space, font=footnotesize, labelfont=bf]{caption}

\usepackage{amsmath,mathrsfs,lineno,amssymb,,amsbsy}

\biboptions{compress}

\journal{}
\makeatother
\newcommand{\HA}[1]{{\color{black} #1}}

\begin{document}

\begin{frontmatter}


\title{The role of frequency and impedance contrasts in bandgap closing and formation patterns of axially-vibrating phononic crystals}

\author[address1,address2]{Hasan B. Al Ba'ba'a}

\author[address2,address3]{Mostafa Nouh\corref{mycorrespondingauthor}}
\cortext[mycorrespondingauthor]{Corresponding author}
\ead{mnouh@buffalo.edu}

\address[address1]{
Department of Mechanical Engineering, Union College, Schenectady, NY 12308, USA}
\address[address2]{
Dept. of Mechanical and Aerospace Engineering, University at Buffalo (SUNY), Buffalo, NY 14260, USA}
\address[address3]{
Dept. of Civil, Structural and Environmental Engineering, University at Buffalo (SUNY), Buffalo, NY 14260, USA}

\begin{abstract}

Bandgaps, or frequency ranges of forbidden wave propagation, are a hallmark of Phononic Crystals (PnCs). Unlike their lattice counterparts, PnCs taking the form of continuous structures exhibit an infinite number of bandgaps of varying location, bandwidth, and distribution along the frequency spectrum. While these bandgaps are commonly predicted from benchmark tools such as the Bloch-wave theory, the conditions that dictate the patterns associated with bandgap symmetry, attenuation, or even closing in multi-bandgap PnCs remain an enigma. In this work, we establish these patterns in one-dimensional rods undergoing longitudinal motion via a canonical transfer-matrix-based approach. In doing so, we connect the conditions governing bandgap formation and closing to their physical origins in the context of the Bragg condition (for infinite media) and natural resonances (for finite counterparts). The developed framework uniquely characterizes individual bandgaps within a larger dispersion spectrum regardless of their parity (i.e., odd vs even bandgaps) or location (low vs high-frequency), by exploiting dimensionless constants of the PnC unit cell which quantify the different contrasts between its constitutive layers. These developments are detailed for a bi-layered PnC and then generalized for a PnC of any number of layers by increasing the model complexity. We envision this mathematical development to be a future standard for the realization of hierarchically-structured PnCs with prescribed and finely tailored bandgap profiles. 

\end{abstract}

\end{frontmatter}

\section{Introduction}

A bandgap, in solid-state physics, is an energy gap in the electronic band structure in which no electronic states exist \cite{holgate2021understanding}. Nearly four decades ago, the birth of photonic crystals gave way to photonic bandgaps, frequency ranges in which all optical modes are absent \cite{john1987strong,yablonovitch1991photonic}. Several years later, phononic crystals---a class of periodic elastoacoustic structures exhibiting forbidden wave propagation within given frequency regimes---extended the definition of bandgaps to the structural dynamics field \cite{sigalas1992elastic}. Ever since, phononic bandgaps have played a central role in several engineering applications ranging from vibroacoustic control \cite{liu2020review} and tunable materials \cite{wang2020tunable}, to topological mechanics \cite{huber2016topological} and nonreciprocal wave phenomena \cite{nassar2020nonreciprocity}.

In its basic form, a phononic crystal (PnC) is a multi-layered composite where the layers self-repeat over an extended spatial domain. Rooted in the origins of periodic structure theory, studies depicting the unique wave propagation properties of PnCs predate the use of the term itself \cite{Mead1971}. The most common one-dimensional PnC configuration involves two alternating materials (or a single material with alternating cross sections) forming a unit cell, often denoted as a diatomic or bi-layered PnC, in which bandgaps arise from Bragg scattering effects at the material (or geometric) interfaces. For an infinite medium, these Bragg bandgaps are a direct function of the structural periodicity and span one or more well-defined frequency ranges which can be predicted by a Bloch-wave analysis of the unit cell \cite{bragg_LR1}. Increasing the number of unit cell layers (or atoms) gives rise to additional features which are uniquely defined by the sequencing and permutations of these individual layers \cite{AlBabaa2019DispersionCrystals}. Bandgap engineering, the science of manipulating phononic parameters within the infinite design space to achieve bandgaps of prescribed characteristics (e.g., bounds, location, attenuation level, targeted modes, directionality, and topological nature, among others) has significantly evolved \cite{oudich2023tailoring}. In pursuit of such goal, studies have utilized geometric properties \cite{yuan2014phononic,li2019effects}, material anisotropy \cite{lin2011tunable,wang2016formation}, damping \cite{tang2020band,bacquet2018metadamping}, viscoelasticity \cite{aladwani2021strategic}, inertance \cite{aladwani2022tunable}, pillared surfaces \cite{jin2021physics}, topology optimization \cite{li2019topology}, and machine learning \cite{oddiraju2022inverse} as tunable knobs in an attempt to tune and achieve maximum control over the bandgap emergence process.

While the applications and utility of these bandgaps in novel and imaginative realizations of PnCs remain an active research area, especially with recent advances in manufacturing and fabrication, the physics underpinning the existence, formation mechanisms, and evolution of phononic bandgaps show intriguing phenomena which continue to be separately explored. Notable among these is the underlying connection between the dispersion relation of an infinite PnC relating the wavenumber of a wave to its frequency, and dictating the frequency-dependent phase and group velocities of a dispersive medium (of which a PnC is one) \cite{kushwaha1993acoustic}, and the structural resonances of a finite PnC where size and boundary effects become intrinsic to the dynamical problem \cite{albabaa2017PC,albabaa2018_Plate}. This interplay between the mathematical description of infinite and finite media, and the ability to recover one from the other \cite{albabaa2017PZ}, was used to develop the theory of truncation resonances in finite PnCs by identifying a set of unique natural frequencies which avert dispersion branches at the infinite limit of the constitutive unit cell \cite{al2023theory,bastawrous2022closed}. Furthermore, understanding the origination process of bandgaps in PnCs and the different ways in which wave attenuation manifests itself in finite periodic media has enabled phononic bandgaps to be artificially emulated in non-periodic lattices \cite{AlBabaa2018BandShaping}, or generated through radically different mechanisms such as inertial amplification \cite{yilmaz2010theory,Orta2019InertialMechanism}. 

Phononic bandgaps are accurately predicted from the conventional Bloch-wave analysis. However, PnCs made of solid continua exhibit a large number of bandgaps which vary in width, strength, and distribution, thus giving rise to the notion of ``bandgap patterns". As this work will show, these patterns are not random and bandgap arrangements in continuous PnCs are far from arbitrary. More importantly, bandgaps that obey certain conditions can be made to vanish (i.e., close by virtue of the preceding and following dispersion branches touching each other), thus rendering the mere existence of such bandgaps in phononic crystals not guaranteed. Instead of deploying numerical tools to seek bandgaps of desirable parameters, this work develops a generalized analytical framework which derives and unravels bandgap patterns and closing conditions in one-dimensional PnC rods undergoing longitudinal motion. This framework is then used to establish general rules which govern bandgap widths and folding frequencies, and connects deformational mode shapes of the culminating PnC to its constitutive layers. In doing so, we explain the conditions driving bandgap closing and connect them to physical origins in the context of the Bragg condition (for infinite media) and natural resonances (for finite counterparts). These developments are detailed for a bi-layered PnC and then generalized for a PnC of any number of layers by increasing the mathematical complexity, while retaining the fully-analytical nature of the model. The need to tailor phononic dispersion profiles have already been shown to play an instrumental role in metamaterial applications \cite{goh2019group, goh2019inverse, morris2022expanding, kazemi2023non}. As such, we envision this mathematical development to be a future standard for designing bandgaps in PnCs with versatile and precisely targeted bandgap profiles.

\begin{figure*}[ht]
     \centering
\includegraphics[]{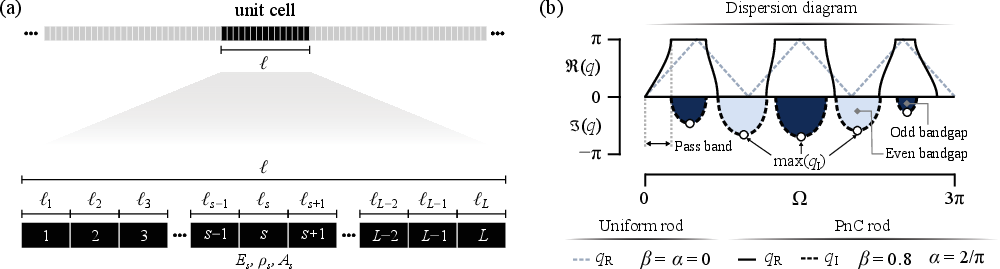}
\caption{(a) Single unit cell of a multi-layered PnC rod of $L$ layers with the geometric and material parameters indicated. (b) Dispersion diagram of a bi-layered ($L=2$) PnC rod with $\alpha = 2/\pi$ and $\beta = 0.8$. The linear dispersion diagram of a uniform rod ($\alpha = \beta = 0$) is also provided for reference. Bandgap regions are shaded and color-coded depending on their parity, i.e., odd versus even.}
     \label{fig:Multi_layer_PnC}
\end{figure*}

\section{Mathematical foundation}
\subsection{PnC configuration}

Starting with the most general case, we consider a continuum PnC in the form of a one-dimensional rod which consists of self-repeating unit cells, where each cell is comprised of $L$ material layers, as depicted in Figure~\ref{fig:Multi_layer_PnC}(a). In this work, we exclude any flexural and torsional waves, and focus on longitudinal motion described by the continuous function $u(x,t)$. In the proposed PnC, each layer has unique geometrical and mechanical properties that do not necessarily match the rest.~The $s^\text{th}$ layer of the unit cell has a mechanical impedance $z_{s} = A_{s}\sqrt{E_{s} \rho_{s}}$ and a sonic speed $c_{s} = \sqrt{E_{s}/\rho_{s}}$, where $E_{s}$, $\rho_{s}$, and $A_{s}$ are the elastic modulus, density, and cross-sectional area, respectively ($s = 1,2, \dots, L$). The lumped parameter (spring-mass) description of this model is commonly referred to as a polyatomic PnC \cite{AlBabaa2019DispersionCrystals}, with each layer within the unit cell denoted as an ``atom". The unit cell's total length is $\ell = \sum_s \ell_s$, which is analogous to the lattice constant of a one-dimensional polyatomic PnC.

\subsection{Transfer matrix method}
The dispersion relation of the aforementioned unit cell can be analytically derived via the transfer matrix method (TMM). The transfer matrix $\mathbf{T}$ obtains the displacement $u$ and force $f$ at the end of cell $i$ from their counterparts at the end of cell $i-1$, such that:
\begin{equation}
    \begin{Bmatrix}
    u_{i} \\ f_{i}
    \end{Bmatrix} = \mathbf{T}
    \begin{Bmatrix}
    u_{i-1} \\ f_{i-1}
    \end{Bmatrix}
    \label{eq:T_matrix}
\end{equation} 
where $\mathbf{T}$ is computed from a series multiplication of the transfer matrices of the individual layers:
\begin{equation}
    \mathbf{T} = \mathbf{T}_L \mathbf{T}_{L-1} \dots \mathbf{T}_{1}
    \label{eq:T_multi_Layer}
\end{equation}
Starting with the one-dimensional wave equation which describes axial waves in the rod, the transfer matrix of the $s^\text{th}$ layer $\mathbf{T}_{s}$ in Eq.~(\ref{eq:T_multi_Layer}) can be derived as \cite{AlBabaa2016a}:
\begin{equation}
    \mathbf{T}_{s} = 
    \begin{bmatrix}
    \cos(k_s \ell_{s}) & \frac{1}{z_s \omega}\sin(k_s \ell_{s}) \\
    - z_s \omega\sin(k_s \ell_{s}) & \cos(k_s \ell_{s})
    \end{bmatrix}
    \label{eq:Ts}
\end{equation}
where $k_{s} = \omega/c_{s}$ denotes the wavenumber within an individual layer, which is a function of the angular frequency~$\omega$.

\section{Bi-layered PnCs}
\subsection{Dispersion Analysis \label{sec:bilayered_disp}}
A special case of the aforementioned structure is the bi-layered PnC rod (i.e., $L = 2$), which will be considered here in detail. In such a case, the transfer matrix of a unit cell $\mathbf{T} = \mathbf{T}_2 \mathbf{T}_1$ is the product of the transfer matrices of the two layers. The resultant $\mathbf{T}$ can be simplified by introducing $\omega_s = c_s/\ell_s$ (where $s=1,2$) and two non-dimensional parameters, namely the frequency and impedance contrasts, respectively, as follows:
\begin{subequations}
\begin{equation}
    \alpha =  \frac{\frac{1}{\omega_1}- \frac{1}{\omega_2}}{\frac{1}{\omega_1}+ \frac{1}{\omega_2} }
    \label{eq:alpha}
\end{equation}
\begin{equation}
    \beta = \frac{z_1 - z_2}{z_1 + z_2} 
    \label{eq:beta}
\end{equation}
\end{subequations}
which both range from $-1$ to $1$ depending on the choice of unit cell parameters. By defining an average impedance of both layers $z = (z_1+z_2)/2$ and using the definition of the impedance contrast $\beta$, the impedance of each layer can be written as:
\begin{equation}
    z_{1,2} = z(1\pm \beta)
\end{equation}
where $+$~($-$) is for the first (second) layer. We also define a non-dimensional frequency $\Omega = \omega/\omega_0$, where the normalization constant $\omega_0$ represents the harmonic mean of $\omega_{1}$ and $\omega_{2}$, and is given by:
\begin{equation}
    \omega_0 = \frac{2}{\frac{1}{\omega_1}+ \frac{1}{\omega_2}}
    \label{eq:harmonic_mean}
\end{equation}
The harmonic mean can be then combined with the definition of $\alpha$ to give:
\begin{equation}
    \frac{1}{\omega_{1,2}} = \frac{1}{\omega_0} (1\pm\alpha)
    \label{eq:Oms}
\end{equation}
where, once again, $+$~($-$) is for the first (second) layer. Using these definitions, the transfer matrix $\mathbf{T}$ of the bi-layered cell can be rewritten as:
\begin{equation}
    \mathbf{T} = 
    \begin{bmatrix}
    d_{-} & \frac{1}{z \omega (1-\beta^2)}t_{-} \\
    -z \omega t_{+} & d_{+}
    \end{bmatrix}
    \label{eq:uc_TM}
\end{equation}
where
\begin{subequations}
\begin{equation}
    d_{\pm} = \frac{1}{1\pm\beta} \Big(\cos(2\Omega) \pm \beta \cos(2\Omega \alpha) \Big)
    \label{eq:dpm}
\end{equation}
\begin{equation}
    t_{\pm} = \sin(2\Omega) \pm \beta \sin(2\Omega \alpha)
    \label{eq:t21}
\end{equation}
\label{eq:UC_TM}
\end{subequations}
Finally, the dispersion relation can be found from $\mathbf{T}$ via $\text{tr}(\mathbf{T}) = 2 \cos(q)$, where $\text{tr}(\mathbf{T})$ is the trace of the matrix $\mathbf{T}$ \HA{(A detailed process outlining the origin of the dispersion relation is provided in \textbf{Appendix A})}. Also, $q = \tilde{q} \ell =  q_\text{R} + \mathbf{i} q_\text{I}$ is the non-dimensional wavenumber of the PnC rod which is a product of the wavenumber $\tilde{q}$ and the unit cell length $\ell$, and $q_\text{R}$ and $q_\text{I}$ denote its real and imaginary components, respectively. Using Eq.~(\ref{eq:dpm}), the dispersion relation is obtained as:
\begin{equation}
    \cos(2 \Omega) - \beta^2 \cos(2 \alpha \Omega) - (1-\beta^2) \cos(q) = 0
    \label{eq:non-dim_disp_rel}
\end{equation}
Note that a positive or negative value of $\alpha$ or $\beta$ does not change the resulting dispersion relation as long as their magnitudes remain the same. This fact can be inferred from the dispersion relation in Eq.~(\ref{eq:non-dim_disp_rel}), where $\alpha$ is in the argument of the even cosine function and $\beta$ is squared. Equation~(\ref{eq:non-dim_disp_rel}) can be depicted analytically by reformulating it as $ q = \cos^{-1} [\Phi(\Omega)]$, where 
\begin{equation}
    \Phi(\Omega) = \frac{1}{(1-\beta^2)} \left[ \cos(2 \Omega) - \beta^2 \cos (2\Omega \alpha) \right]
    \label{eq:Phi}
\end{equation}
An interesting feature of the function $\Phi(\Omega)$ is its association with the frequencies of maximum attenuation inside Bragg bandgaps, which can be found by evaluating the roots of the derivative $\partial \Phi(\Omega)/\partial \Omega = 0$, analogous to lumped PnCs \cite{AlBabaa2019DispersionCrystals,albabaa2017PC}, yielding:
\begin{equation}
    \sin(2\Omega) - \beta^2 \alpha \sin(2\alpha \Omega) = 0
    \label{eq:Max_Att}
\end{equation}

\vspace{0.1cm}

Figure~\ref{fig:Multi_layer_PnC}(b) shows two dispersion diagrams for a bi-layered PnC rod (with $\alpha = 2/\pi$ and $\beta = 0.8$) and a uniform rod with two identical layers (i.e., $\alpha = \beta = 0$). The uniform rod exhibit linear dispersion, a hallmark feature of longitudinal elastic waves in rods \cite{Hussein2014}, while the PnC's dispersion relation is nonlinear culminating in a dispersive behavior.~The bandgaps in the PnC case align with $q_\text{I} \neq 0$ regions.~Odd and even-numbered bandgaps are color-coded for easier interpretation. Their analytical derivation and emergence conditions are established next.

\begin{figure*}[ht]
     \centering
\includegraphics[]{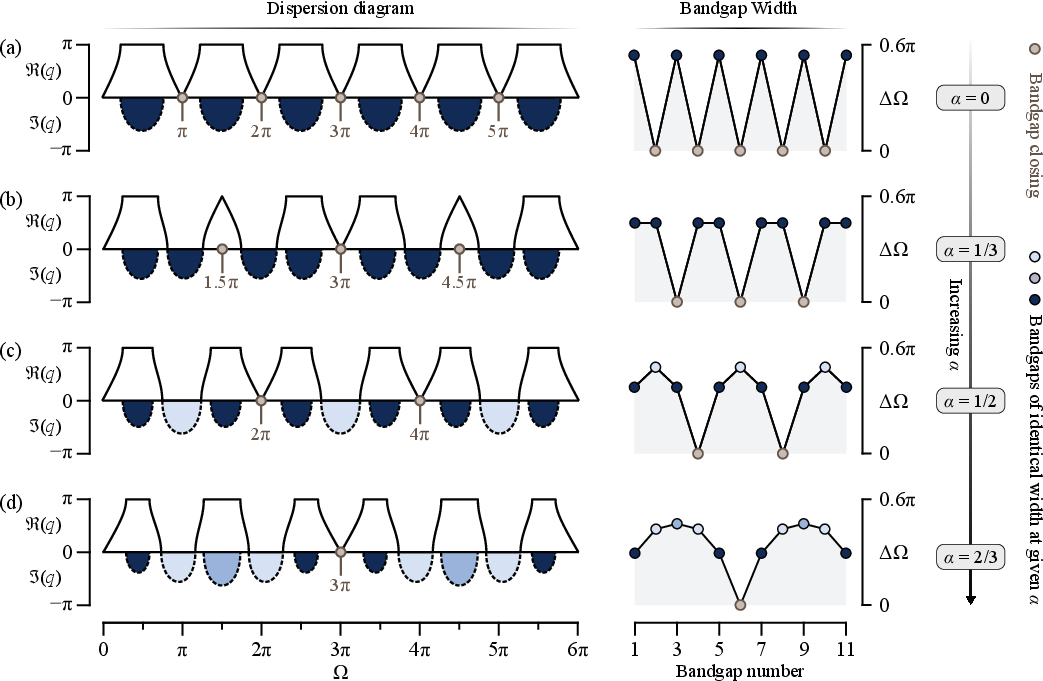}
     \caption{Dispersion diagram (left) and the corresponding width ($\Delta \Omega$) of the first 11 bandgaps (right) for a bi-layered PnC rod with (a) $\alpha = 0$, (b) $\alpha = 1/3$, (c) $\alpha = 1/2$, and (d) $\alpha = 2/3$. $\beta = -0.75$ is used for all cases. Regions of the same shading color indicate bandgaps of identical width at a given $\alpha$, while the labeled frequencies (e.g., $\pi$, $2\pi$, etc.) denote bandgap closings.}
     \label{fig:alpha_disp_examples}
\end{figure*}
\begin{figure*}[h]
     \centering
\includegraphics[]{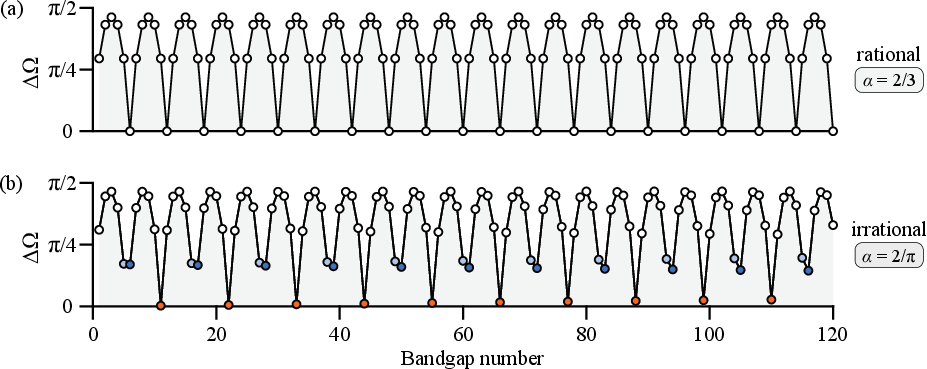}
\caption{Bandgap width for the first 120 bandgaps of a bi-layered PnC rod with: (a) a rational value of $\alpha=2/3$ and (b) an irrational value of $\alpha=2/\pi$. $\beta = -0.75$ is used for both cases. The PnC with the rational $\alpha$ maintains a perfectly periodic pattern of bandgap widths that repeats itself every 6 bandgaps. The PnC with the irrational $\alpha$ has an aperiodic profile of bandgap widths which can be observed by tracking the changes in the widths of the 5$^{\text{th}}$ (light blue marker), 6$^{\text{th}}$ (dark blue marker), and 11$^{\text{th}}$ (orange marker) every cycle of 11 bandgaps.}
\label{fig:irrational_alpha}
\end{figure*}

\subsection{Bandgap closing in bi-layered PnCs}
\label{sec:BG-conditions}
Bandgap limits for odd and even-numbered bandgaps can be found by substituting $q = \pi$ and $q = 0$, respectively, in the dispersion relation shown in Eq.~(\ref{eq:non-dim_disp_rel}), resulting in the following expressions:
\begin{subequations}
\begin{equation}
    \big(\cos(\Omega) - \beta \cos(\alpha \Omega) \big)\big(\cos(\Omega) + \beta \cos(\alpha \Omega) \big) = 0
\end{equation}
\begin{equation}
    \big(\sin(\Omega) - \beta \sin(\alpha \Omega) \big)\big(\sin(\Omega) + \beta \sin(\alpha \Omega) \big) = 0
\end{equation}
\end{subequations}
which amount to a multiplication of two terms, each of which giving one bandgap limit. It is also evident that a bandgap only emerges if the roots of each of these terms are different. Thus, by equating both terms in each equation, the conditions that render a bandgap width equal to zero can be obtained (commonly referred to as zero-width bandgaps \cite{hvatov2015free}). These conditions are summarized by the following equations:
\begin{subequations}
\begin{equation}
    \beta \cos(\alpha \Omega) = 0
\end{equation}
\begin{equation}
    \beta \sin(\alpha \Omega) = 0
\end{equation}
\label{eq:odd-even-BG-closing-limits}
\end{subequations}
for odd and even-numbered bandgaps, respectively. Consider the cases that satisfy Eq.~(\ref{eq:odd-even-BG-closing-limits}) when $\beta\neq0$. Starting with odd-numbered bandgaps, a zero-width bandgap needs to satisfy $\cos(\alpha \Omega)=0$, a condition which is guaranteed at the following frequencies:
\begin{equation}
    \Omega = \frac{(2r-1)\pi}{2\alpha}; \ \ \HA{\text{for} \ \alpha \neq 0}
    \label{eq:Om_odd_BG_closing_2}
\end{equation}
for $r \in \mathbb{N}^+$, where $\mathbb{N}^+$ are all natural numbers without zero. The second requirement is that such frequencies in Eq.~(\ref{eq:Om_odd_BG_closing_2}) have to satisfy the dispersion relation at $q = \pi$, which can be checked by plugging Eq.~(\ref{eq:Om_odd_BG_closing_2}) in (\ref{eq:non-dim_disp_rel}), and setting $q = \pi$, which, after simplification, gives:
\begin{equation}
    \cos\left(\frac{\pi}{\alpha}(2r-1)\right) = -1
\end{equation}
Solving for $\alpha$, we arrive at the following expression:
\begin{equation}
    \alpha = \frac{\alpha_n}{\alpha_d} = \frac{(2r-1)}{(2p-1)}
    \label{eq:alpha_rational}
\end{equation}
for $p\in \mathbb{N}^+$, indicating that some odd-numbered bandgaps close when $\alpha$ is a rational number of odd integers. The frequencies at which the bandgap closes can be found by combining Eqs.~(\ref{eq:alpha_rational}) and (\ref{eq:Om_odd_BG_closing_2}), yielding:
\begin{equation}
    \Omega_p = \frac{\pi}{2}(2p-1)
    \label{eq:om_p_odd_1}
\end{equation}
\HA{It should be noted, however, that if a specific ratio of $\alpha = \alpha_n/\alpha_d$ is imposed, only select combinations of $r$ and $p$ will maintain such ratio. In this scenario, not every $p\in \mathbb{N}^+$ is guaranteed to fulfill this ratio of $\alpha$, and Eq.~(\ref{eq:om_p_odd_1}) needs to be updated to:
\begin{equation}
    \Omega_p = \frac{\pi}{2}\alpha_d (2p-1); \ \ \text{for} \ p \in \mathbb{N}^+
    \label{eq:om_p_odd}
\end{equation}
to guarantee bandgap closing. To illustrate, consider the case where $\alpha = 1/3$ (i.e., $\alpha_d = 3$). A combination of $r=1$ and $p=2$ therefore closes one bandgap at $\Omega_p = 3\pi/2$, while a combination of $r=2$ and $p=5$ closes another bandgap at $\Omega_p = 9\pi/2$. However, even though $p=3$ belongs to $p\in \mathbb{N}^+$, there exists no $r\in \mathbb{N}^+$ that satisfies the chosen $\alpha$. As a result, plugging $p=3$ into Eq.~(\ref{eq:om_p_odd_1}) would \textbf{not} result in a bandgap closing frequency for the PnC described by this $\alpha$, making it prudent to use Eq.~(\ref{eq:om_p_odd}) instead.}

\vspace{0.1cm}

Similarly, we analyze even-numbered bandgaps to find $\alpha$ values at which they vanish. Knowing that $\sin(\alpha \Omega) = 0$ must be met for such a case, we have $\alpha = 0$ and
\begin{equation}
    \Omega = \frac{r\pi}{\alpha}; \ \ \HA{\text{for} \ \alpha \neq 0}
    \label{eq:omega_even_numbered_BGs}
\end{equation}
Plugging Eq.~(\ref{eq:omega_even_numbered_BGs}) back in (\ref{eq:non-dim_disp_rel}), and setting $q = 0$, the values of $\alpha$ that correspond to zero-width even-numbered bandgaps are also rational and given by:
\begin{equation}
    \alpha = \frac{\alpha_n}{\alpha_d} = \frac{r}{p}
\end{equation}
It should be noted that all rational values of $\alpha$ will close even-numbered bandgaps, occurring at the following frequencies:
\begin{equation}
    \Omega_{p} = \pi p \alpha_d; \ \ \text{for} \ p \in \mathbb{N}^+
    \label{eq:om_p_even}
\end{equation}

Figure~\ref{fig:alpha_disp_examples} show examples of dispersion relations with $\beta = -0.75$ and different values of $\alpha$, namely, (a) $\alpha = 0$, (b) $\alpha = 1/3$, (c) $\alpha = 1/2$, and (d) $\alpha = 2/3$. The left panel of the figure displays the full dispersion diagrams, while the right panel graphically represents the bandgap width of all eleven bandgaps in the range $\Omega \in [0,6\pi]$. In all of the shown cases, even-numbered bandgaps that occur at the frequencies described in Eq.~(\ref{eq:om_p_even}) are expected to close for all $\alpha$ values that are rational. The first case of $\alpha = 0$, however, is a special case where all even-numbered bandgaps close, while all odd-numbered bandgaps remain open and maintain identical widths, as can be inferred from the right subplot of Figure~\ref{fig:alpha_disp_examples}(a). \HA{Note that for $\alpha = 0$ and $\beta \neq 0$, the maximum attenuation can be found in closed-form at the frequencies $\Omega =r\pi/2$ (See \textbf{Appendix B} for more details on the $\alpha=0$ case).} For the second case of $\alpha = 1/3$, the numerator and denominator of $\alpha$ are odd integers. It is therefore expected to see odd and even-numbered bandgaps calculated from Eqs.~(\ref{eq:om_p_odd}) and (\ref{eq:om_p_even}) closed. The closing takes place at multiples of $\Omega = 1.5\pi$. The third and fourth cases of $\alpha = 1/2$ and $\alpha = 2/3$, respectively, have numerator and denominator values with different parities. As a result, only even-numbered bandgaps are expected to close, which takes place at $\Omega = 2\pi,4\pi$ and $\Omega = 3\pi$, respectively. 

\vspace{0.1cm}

As can be observed from the right panel of Figure~\ref{fig:alpha_disp_examples}, bandgap width profiles exhibit a wave-like behavior for all considered values of $\alpha$, which perfectly repeats itself. Additionally, these profiles are noted to be mirror-symmetric around the closing points. Finally, we emphasize that the order of the bandgaps that close are always related to $\alpha$, except for the special case of $\alpha = 0$. Specifically, the order of closed bandgaps is equal to multiples of $\alpha_d$ if both numerator and denominator are odd, while equal to twice the multiples of $\alpha_d$ otherwise. 

\subsection{Rational versus irrational $\alpha$ values}
\label{sec:irrational-alpha}

Following this discussion of the role of rational $\alpha$ values in bandgap closure, it is imperative to understand the different consequences of PnCs with rational and irrational $\alpha$ values that are close in magnitude. \HA{Consider two bi-layered unit cells of an identical impedance contrast $\beta = -0.75$ with $\alpha = 2/3$ for the first, which is the rational value that corresponds to the dispersion diagram shown in Figure~\ref{fig:alpha_disp_examples}(d), and $\alpha = 2/\pi$ for the second, which is the irrational value used to construct the dispersion diagram shown in Figure~\ref{fig:Multi_layer_PnC}(b). The bandgap widths for the first 120 bandgaps of both PnCs is computed in Figure~\ref{fig:irrational_alpha}. It is immediately noticed that the rational ($\alpha = 2/3$) case maintains a perfectly periodic pattern of bandgap widths that repeats itself every 6 bandgaps (which is twice $\alpha_d$ as explained earlier). On the other hand, the bandgap widths corresponding to the irrational ($\alpha = 2/\pi$) case clearly move further away from zero as the bandgap number increases, indicating the absence of a bandgap closing pattern due to $\alpha$ not being an exact rational number. Despite the absence of a bandgap closing pattern, Figure~\ref{fig:irrational_alpha}(b) still shows a near-periodic profile with a period of 11 bandgaps. This is understandeable because the closest rational approximation of $\alpha = 2/\pi \approx 2/(22/7) \approx 7/11$ (using the known approximation of $\pi$) reveals that this system should closely mimic one which exhibits bandgap closing at multiples of $\alpha_d = 11$.}

\subsection{Physical implication of rational $\alpha$ values}

Consider a finite uniform rod made of one of the two bi-layered PnC unit cell layers, that has a sonic speed $c_s$ and a length $\ell_s$. Excluding the rigid body mode at $\omega=0$, the natural frequencies of this rod in the unconstrained form (i.e., with free-free boundary conditions) can be found by setting the lower off-diagonal element of the transfer matrix in Eq.~(\ref{eq:Ts}) equal to zero. If the same rod is fixed from both ends, the natural frequencies can be obtained by setting the upper off-diagonal element of the transfer matrix in Eq.~(\ref{eq:Ts}) equal to zero. Both cases yield $\sin(k_s \ell_s) = 0$, which in turn provides the following set of natural frequencies \cite{al2023theory}: 
\begin{equation}
    \omega = \pi n_s \omega_s; \hspace{0.5 cm} n_s = 1,2,3,\dots
    \label{eq:nat_freq_rods}
\end{equation}
where $n_s$ indicates the order of the vibrational mode in the complete set of non-zero natural frequencies.~Using Eq.~(\ref{eq:Oms}) and the substitution $\alpha = \alpha_n/\alpha_d$, combined with Eq.~(\ref{eq:nat_freq_rods}), we arrive at:
\begin{equation}
(\alpha_d - \alpha_n) n_1
= (\alpha_d + \alpha_n) n_2
\label{eq:harmonics_equal}
\end{equation}

Equation~(\ref{eq:harmonics_equal}) captures the physical meaning behind rational values of $\alpha$ in a concise manner. It indicates that for any rational $\alpha$ value, a natural frequency of the first of the two layers of the order $(\alpha_d - \alpha_n)n_1$ matches a natural frequency of the second layer of the order $(\alpha_d + \alpha_n)n_2$, since $\alpha_d$ and $\alpha_n$ are integers.~These natural frequencies must satisfy $\Omega_p$ in Eqs.~(\ref{eq:om_p_odd}) and (\ref{eq:om_p_even}). Consequently, along with the harmonic mean in Eq.~(\ref{eq:harmonic_mean}), these two equations can be utilized to find an exact solution for $n_1$ and $n_2$ for a prescribed rational value of $\alpha$. Rearranging Eq.~(\ref{eq:harmonic_mean}), it can be seen that:
\begin{equation}
    \frac{\omega_0}{\omega_1} + \frac{\omega_0}{\omega_2} = 2
    \label{eq:ns_2}
\end{equation}
which, in conjunction with Eq.~(\ref{eq:nat_freq_rods}) at $\omega = \omega_0 \Omega_p$, becomes:
\begin{equation}
   \pi(n_1 + n_2) = 2\Omega_p
    \label{eq:ns_3}
\end{equation}
Solving Eqs.~(\ref{eq:harmonics_equal}) and (\ref{eq:ns_3}) simultaneously and plugging in the expressions for $\Omega_p$ in Eqs.~(\ref{eq:om_p_odd}) and (\ref{eq:om_p_even}), we arrive at the following expressions for $n_{1,2}$:
\begin{subequations}
\begin{equation}
    n_s = \frac{1}{2}(2p-1)(\alpha_d\pm\alpha_n)
\end{equation}
\begin{equation}
    n_s = p(\alpha_d\pm\alpha_n)
\end{equation}
\label{eq:n_s_integer_solution}
\end{subequations}
for odd and even-numbered bandgaps, respectively, and with $+$~($-$) denoting the solution for $s=1$ ($s=2$). Note that if the sign of $\alpha$ flips, the solutions corresponding to the first PnC layer become those of the second layer and vice versa.~Interestingly, this discussion of the physical meaning of rational $\alpha$ values also has a connection to the ``Bragg condition", as will be derived next. 

\subsection{Connection to Bragg condition}

Bandgaps in PnCs are known to be size-dependent and initiate near frequencies described by the Bragg condition, which provides the proportional relationship between the size of a PnC and the wavelength \cite{xiao2012longitudinal, Xiao2012FlexuralResonators}. A Bragg condition can also be defined for each of the individual layers of a bi-layered PnC (since the individual layer can be thought of as a special PnC unit cell with zero-width bandgaps) as:
\begin{equation}
    \ell_s = n_s\frac{\lambda_s}{2}
    \label{eq:Bragg-condition}
\end{equation}
where $\lambda_s = 2 \pi/{k_s}$ is the wavelength.~Recalling that $k_s=\omega/c_s$ for rods and rearranging Eq.~(\ref{eq:Bragg-condition}) in terms of the angular frequency $\omega$, it can be seen that the frequencies corresponding to the Bragg condition are given by $\omega = \pi n_s \omega_s$, which perfectly align with the frequencies derived in Eq.~(\ref{eq:nat_freq_rods}). In other words, the frequencies corresponding to the Bragg condition of an individual layer are also the natural frequencies of a finite uniform rod comprised of that particular layer with free-free or fixed-fixed boundaries. By making use of Eq.~(\ref{eq:Oms}), a non-dimensional form of the Bragg condition of each individual layer in a bi-layered PnC can be written as a function of the frequency contrast between the two layers $\alpha$, as follows:
\begin{equation}
    \Omega =\frac{ n_{s} \pi }{ 1\pm \alpha}
    \label{eq:Bragg-cond-pm-alpha}
\end{equation}
where $+$~($-$) is for $s=1$ ($s=2$). Recalling that a bandgap of a bi-layered PnC can only close if $\alpha$ is a rational number (as proven in Section~\ref{sec:BG-conditions}), it can be seen that a rational $\alpha$ is guaranteed if the frequencies of the Bragg condition for the individual layers 1 and 2 are matched, i.e., when $\Omega$ in Eq.~(\ref{eq:Bragg-cond-pm-alpha}) becomes identical for both the plus and minus solutions. As a result of this matching, $\alpha$ takes the following expression: 
\begin{equation}
    \alpha = \frac{n_1 - n_2}{n_1 + n_2}
    \label{eq:alpha-Bragg}
\end{equation}
thus ensuring that $\alpha$ is a rational number, and further cementing the connection between the Bragg condition and the bandgap closing condition in a bi-layered PnC rod. 

\subsection{Mode shapes at bandgap closing}

\begin{figure*}[ht]
     \centering
\includegraphics[]{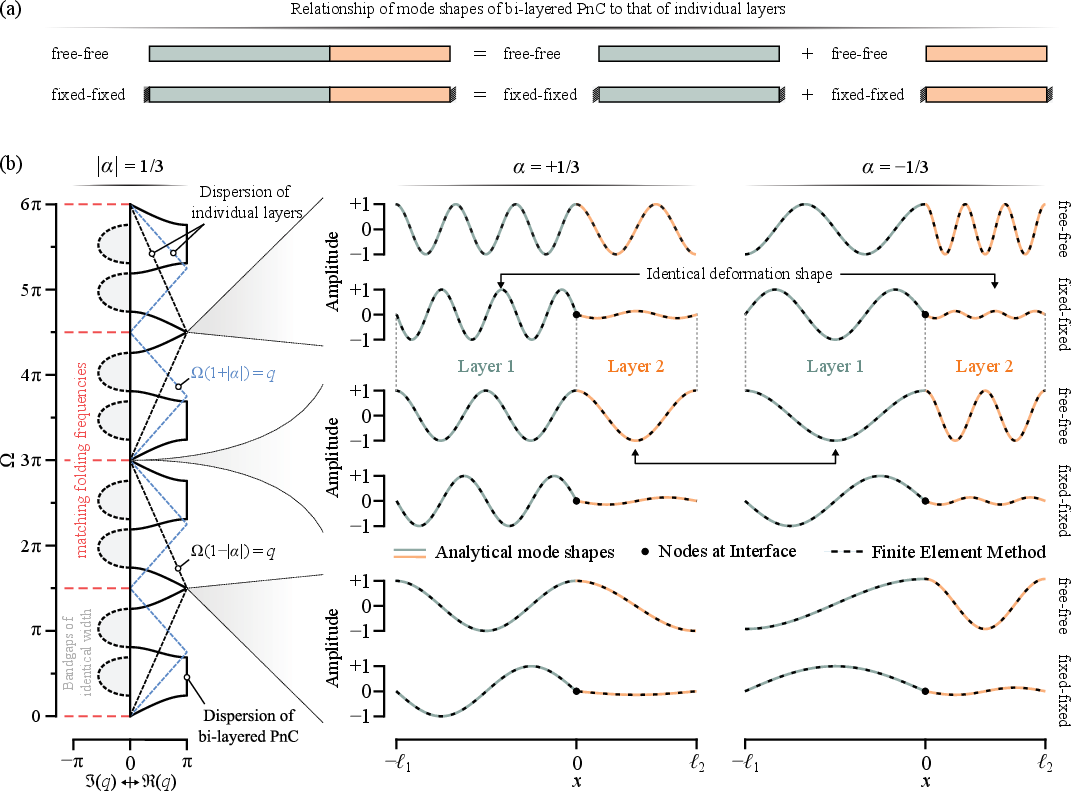}
     \caption{(a) A graphical summary of the relationship of the mode shapes (for given boundary conditions) of a PnC unit cell to that of its individual layers at bandgap closing frequencies. (b) \textit{Left:} Dispersion diagram of a bi-layered PnC rod unit cell for the case when $|\alpha| = 1/3$ and $\beta = -0.75$. Two sets of folded lines represent the dispersion diagrams of the two individual layers of the PnC. Red dashed lines indicate the locations at which the two sets fold at the same frequency, indicating a bandgap closing of the bi-layered PnC as shown. \textit{Right:} Mode shapes of a bi-layered PnC rod unit cell using the same parameters. The mode shapes are shown for the three frequencies which correspond to bandgap closings within the range $\Omega \in (0,6\pi)$, namely $3\pi/2$, $3\pi$, and $9\pi/2$. Repeated modes exist at the bandgap closings due to two dispersion branches touching at that point. Changing the sign of $\alpha$ flips the modes shape as illustrated in the right panel of the figure. Specifically, the deformation shape of layer 1 at $\alpha = 1/3$ becomes that of layer 2 at $\alpha = -1/3$, and vice versa. Mode shapes calculated via the finite element method are superimposed as dashed lines for validation.}
     \label{fig:mode_shapes}
\end{figure*}

It is now established that the Bragg bandgaps of a bi-layered PnC close when the Bragg condition frequencies of the two constitutive layers match. As a direct consequence of that condition, the natural frequencies of the bi-layered PnC unit cell become those of the individual layers at bandgap closing. We therefore formulate analytical expressions for the deformational mode shapes of the bi-layered PnC (often referred to as the unit cell Bloch modes \cite{ragonese2021prediction}) which correspond to bandgap closing frequencies, for a complete understanding of these scenarios. The general solution of the displacement and internal force of the $s^\text{th}$ layer of the PnC rod can be written as:
\begin{subequations}
\begin{equation}
    u_s(x) = a_s \cos(k_s x) + b_s \sin(k_s x)
\end{equation}
\begin{equation}
    f_s(x) = E_s A_s k_s \big( b_s \cos(k_s x) - a_s \sin(k_s x) \big)
\end{equation}
\end{subequations}
To obtain solutions for the coefficients $a_s$ and $b_s$, a total of four equations are needed which are found from the displacement and force continuity conditions between the PnC layers. To do so in a bi-layered PnC, we set $x=0$ at the interface of the two layers. The displacement and force continuity conditions yield:
\begin{subequations}
\begin{equation}
    a_1 - a_2 = 0
\end{equation}
\begin{equation}
    (1+\beta)b_1 - (1-\beta) b_2 = 0
    \label{eq:b_1_b_2}
\end{equation}
\label{eq:set_1_mode_shapes}
\end{subequations}
By using the next interface to get the two remaining equations, we get $u_1(\ell - \ell_1) = u_2(\ell_2)$ and $f_1(\ell - \ell_1) = f_2(\ell_2)$, which can be expressed as $u_1(\ell - \ell_1) = \text{e}^{\mathbf{i}q}u_1(-\ell_1)$ and $f_1(\ell - \ell_1) = \text{e}^{\mathbf{i}q}f_1(-\ell_1)$, respectively, by virtue of Bloch's theorem. At the special cases of $\sin(k_s \ell_s) = 0$, these two continuity conditions simplify to:
\begin{subequations}
\begin{equation}
    a_1 \cos(n_1\pi) \text{e}^{\mathbf{i}q} - a_2 \cos(n_2\pi) = 0
\end{equation}
\begin{equation}
    (1+\beta) \cos(n_1\pi) \text{e}^{\mathbf{i}q} b_1 - (1-\beta) \cos(n_2\pi) b_2 = 0
\end{equation}
\label{eq:set_2_mode_shapes}
\end{subequations}
Solving Eqs.~(\ref{eq:set_1_mode_shapes})~and~(\ref{eq:set_2_mode_shapes}) simultaneously gives:
\begin{subequations}
\begin{equation}
    \big(\cos(n_1\pi) \text{e}^{\textbf{i}q} - \cos(n_2\pi)\big)a_1 = 0
\end{equation}
\begin{equation}
    \big(\cos(n_1\pi) \text{e}^{\textbf{i}q} - \cos(n_2\pi)\big)b_1 = 0
\end{equation}
\label{eq: eigenvalueprob}
\end{subequations}
Equation~(\ref{eq: eigenvalueprob}) represents an eigenvalue problem with the eigenvectors being $\{a_1 \ b_1\}^{\text{T}} = \{1 \ 0\}^{\text{T}}$ and $\{a_1 \ b_1\}^{\text{T}} = \{0 \ 1\}^{\text{T}}$.~As such, we arrive at two different mode shape equations.~The first corresponds to $a_1 = a_2 = 1$ and $b_1 = b_2 = 0$, i.e.,
\begin{subequations}
\begin{align}
    u_1 (x) = \cos\left(n_1 \pi \frac{x}{\ell_1} \right); & \hspace{0.5 cm} x \in [-\ell_1, 0] 
    \label{eq:mode1} \\
    u_2 (x) = \cos\left(n_2 \pi \frac{x}{\ell_2} \right); & \hspace{0.5 cm} x \in (0, \ell_2]
    \label{eq:mode2}
\end{align}
\label{eq:free-free-mode-shapes}
\end{subequations}
which, interestingly, is independent of the impedance contrast $\beta$ between the two PnC layers. The mode shapes in Eq.~(\ref{eq:free-free-mode-shapes}) mandate that the normalized amplitude at the interface is equal to one. Similarly, a second solution is found by using Eq.~(\ref{eq:b_1_b_2}) and applying $b_1 = 1-\beta$ and $a_1 = a_2 = 0$, resulting in the following mode shape:
\begin{subequations}
\begin{align}
    u_1 (x) = (1-\beta) \sin\left(n_1 \pi \frac{x}{\ell_1} \right); & \hspace{0.5 cm}  x \in [-\ell_1 , 0] \\
    u_2 (x) = (1+\beta) \sin\left(n_2 \pi \frac{x}{\ell_2} \right); & \hspace{0.5 cm} x \in (0, \ell_2] 
\end{align}
\label{eq:fixed-fixed-mode-shapes}
\end{subequations}
showing that mode shapes from this second solution exhibit an amplitude of zero at the interface between the layers. Upon examining the modes shapes in Eqs.~(\ref{eq:free-free-mode-shapes})~and~(\ref{eq:fixed-fixed-mode-shapes}), it becomes clear that the deformation ``shape" of each layer is independent of the other as inferred from the argument of the cosine and sine functions in both equations. The ``amplitude", however, of mode shapes obtained from Eq.~(\ref{eq:fixed-fixed-mode-shapes}) depends on the impedance contrast between the two layers, as implied by the $(1 \mp \beta)$ coefficient. The independent deformation shapes in each layer are attributed to the fact that Eq.~(\ref{eq:free-free-mode-shapes}) is merely a combination of free-free mode shapes for layers 1 and 2 if they are to be stand-alone structures. Likewise, Eq.~(\ref{eq:fixed-fixed-mode-shapes}) describes fixed-fixed mode shapes for layers 1 and 2, only scaled by the $(1\mp \beta)$ term.~This intriguing relationship between the mode shapes of a PnC and those of its constitutive layers at bandgap closing is graphically summarized in Figure~\ref{fig:mode_shapes}(a).

The rightmost panel of Figure~\ref{fig:mode_shapes}(b) shows the mode shapes for the case of $|\alpha| = 1/3$ and $\beta = -0.75$. Here, we chose $\ell_1/\ell = 0.6$ and thus $\ell_2/\ell = 0.4$. The two modes derived earlier are normalized such that the maximum amplitude is unity and they are plotted at the three frequencies corresponding to bandgap closings within the range $\Omega \in (0,6\pi)$, namely $3\pi/2$, $3\pi$, and $9\pi/2$. The spatial frequency of the mode shapes is controlled by the value of $n_{1,2}$. As can be inferred from Eq.~(\ref{eq:n_s_integer_solution}), flipping the sign of $\alpha$ switches the values of $n_1$ and $n_2$.~This is graphically shown in Figure~\ref{fig:mode_shapes}(b) where the deformation shape of layer 1 at $\alpha = 1/3$ becomes identical to that of layer 2 at $\alpha = -1/3$, and vice versa, at any of the three bandgap closing frequencies shown (Understandably, the deformation shape spans a shorter or larger distance when the sign of $\alpha$ is swapped to accommodate for the different lengths of the individual layers). For validation, all of the analytically-obtained results shown in Figure~\ref{fig:mode_shapes}(b) are verified via a finite element model (FEM) implementing two-node rod elements \cite{petyt2010introduction}, and shown as dashed lines in all the plotted mode shapes. 

The leftmost panel of Figure~\ref{fig:mode_shapes}(b) shows several unique features of the dispersion diagrams of the bi-layered PnC and its two constitutive layers. The latter are given by two sets of folded lines described by $\Omega (1\pm \alpha) = q$ (black and blue dashed lines).~The red dashed lines indicate the locations at which the two sets fold at the same frequency, indicating a bandgap closing of the bi-layered PnC as shown.~Finally, it can also be shown that $n_{1,2}$ indicate precisely the number of folded lines that the dispersion relations of the individual layers have up to each bandgap closing of the bi-layered unit cell. For example, consider the first bandgap closing at $\Omega = 3\pi/2$.~The dispersion relation $\Omega(1+\alpha) = q$ before and up to that frequency consists of exactly two folded lines, while $\Omega(1-\alpha) = q$ consists of one folded line, indicating values of $n_1  = 2$ and $n_2 = 1$.~Using these values of $n_{1,2}$, it immediately follows that $\alpha = 1/3$ by using Eq.~(\ref{eq:alpha-Bragg}), as expected.

\begin{figure*}[]
     \centering
\includegraphics[]{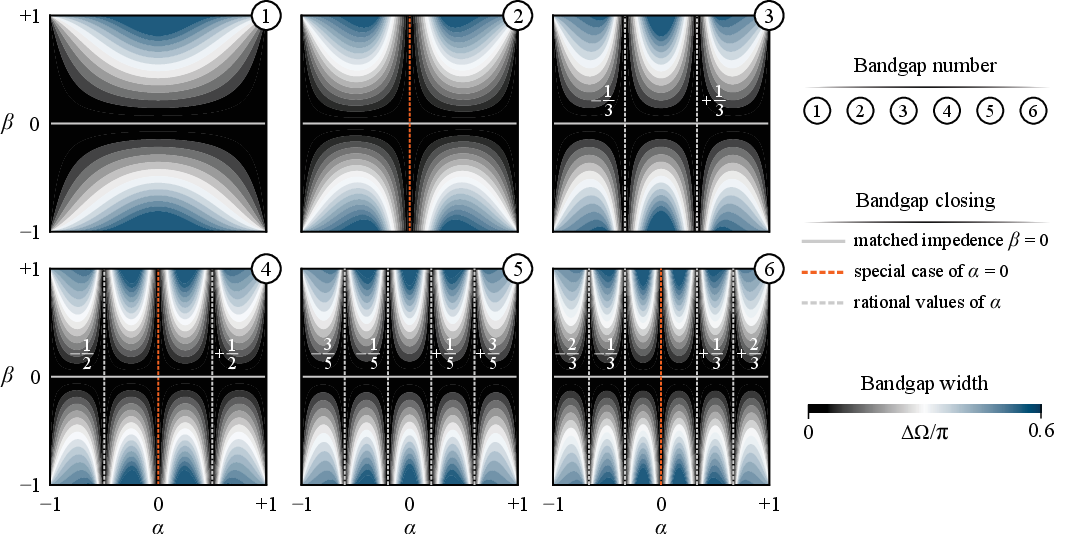}
     \caption{Contours depicting the width $\Delta \Omega$ of the first six bandgaps of a bi-layered PnC rod across the entire design space of $\alpha$ and $\beta$. Bandgap numbers 1 through 6 are placed at the right top corner of each subplot. As can be seen, $\beta = 0$ closes all bandgaps regardless of the value of $\alpha$. $\alpha = 0$ closes all even-numbered bandgaps. Additionally, zero-width bandgaps occur at select rational values of $\alpha$ indicated by the vertical dashed lines, the location of which depends on the bandgap number.}
     \label{fig:alpha_beta_sweep}
\end{figure*}


\subsection{Bandgap transitions with varying $\alpha$ and $\beta$}

The observations drawn in sections~\ref{sec:BG-conditions}~and~\ref{sec:irrational-alpha} regarding rational and irrational values of $\alpha$ are in fact independent of the chosen value of $\beta$. As a demonstration, Figure~\ref{fig:alpha_beta_sweep} shows the width $\Delta \Omega$ of the first six bandgaps over the entire range of $\alpha$ and $\beta$ values. The following observations can be made:

\begin{enumerate}
\itemsep 0.1cm
    \item Confirming the bandgap closing rules observed in Figure~\ref{fig:alpha_disp_examples}, the number of the zero-width bandgap is directly related to the value of $\alpha$.~For example, the fourth bandgap (which is even-numbered) closes at $\alpha = \pm 1/2$ as expected, with the closed bandgap number being twice the denominator value ($\alpha_d = 2$). Similarly, the fifth bandgap (which is odd-numbered) closes at $\alpha = \pm 1/5$ and $\alpha = \pm 3/5$, which are both ratios of odd numbers, and with the closed bandgap number matching the denominator value ($\alpha_d = 5$). Finally, the sixth bandgap closes at both $\alpha = \pm 1/3$ and $\alpha = \pm 2/3$, which represent rational values of identical and different numerator-denominator parity, respectively. 
    
    \item Including the limiting case of $|\alpha| = 1$, the number of times a given bandgap closes is equal to its number plus one. For instance, the third bandgap closes four times at $\alpha = \pm 1/3$ and $\alpha = \pm 1$.
    
    \item The special case of $\alpha = 0$ results in the closing of all even-numbered bandgaps, further confirming the result of Figure~\ref{fig:alpha_disp_examples}(a).
    
    \item The special case of $\beta = 0$ forces all bandgaps to close regardless of the value of $\alpha$ (as reported in \cite{hvatov2015free}).
\end{enumerate}

\begin{figure*}[]
     \centering
\includegraphics[]{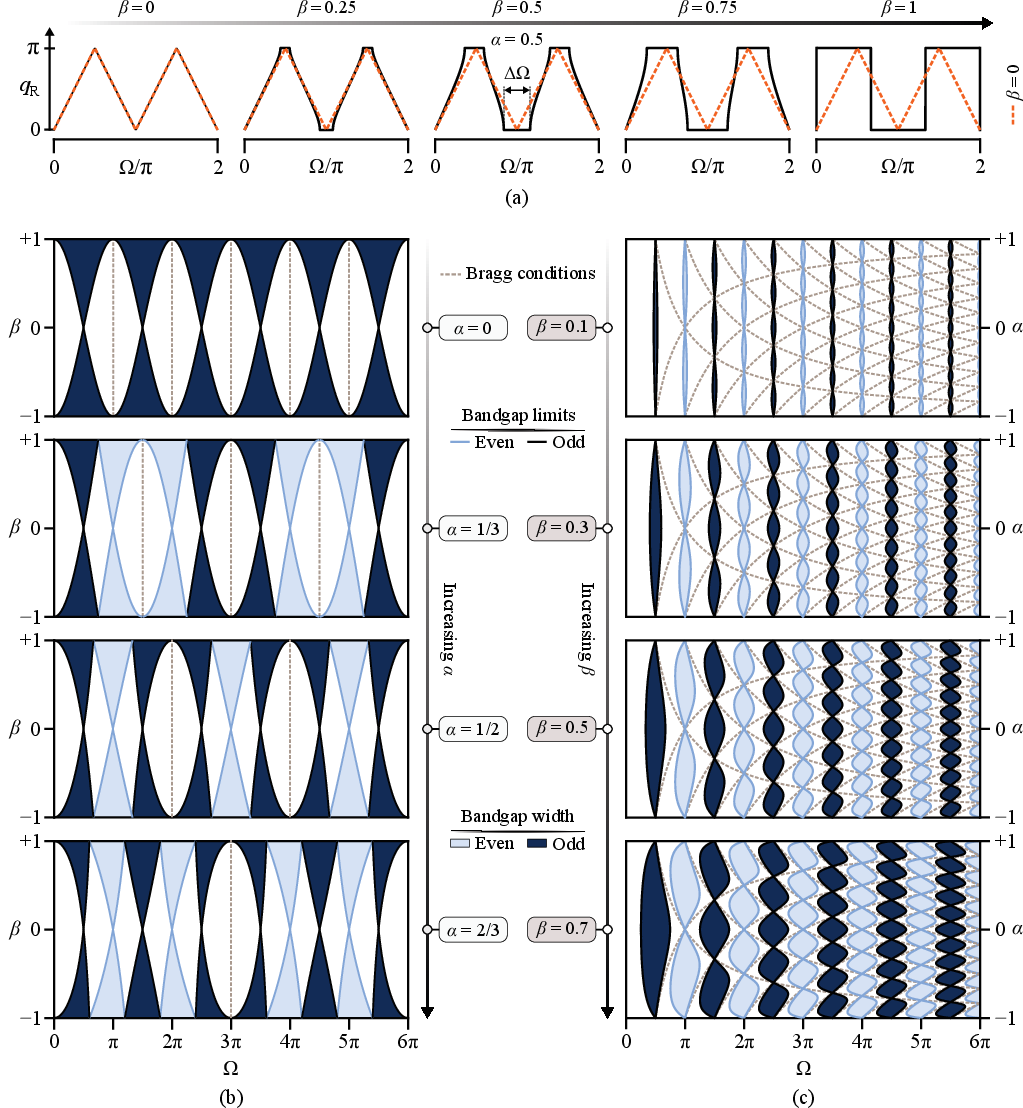}
     \caption{(a) Dispersion diagram of a bi-layered PnC rod with $\alpha =1/2$ and the two layers having varying impedance contrast ranging from $\beta = 0$ to $\beta = 1$. The $\beta = 0$ case, which is shown as dashed orange lines in all plots for reference, represents a bi-layered PnC unit cell with zero impedance contrast between its two layers (i.e., two layers with the same impedance) and shows no bandgap emergence (i.e., $\Delta \Omega = 0$). As $\beta$ increases, bandgaps initiate at the folding frequencies and the bandgap width $\Delta \Omega$ continues to grow until the dispersion branches become completely flat at $\beta = 1$. Evolution of bandgap limits of a bi-layered PnC rod for: (b) varying $\beta$ at specific $\alpha$ values of $0, 1/3, 1/2, \text{and} \ 2/3$ , and (c) varying $\alpha$ at specific $\beta$ values of $0.1, 0.3, 0.5, \text{and} \ 0.7$. In (b), the dashed lines represent frequencies where bandgap limits match the Bragg conditions derived in Eq.~(\ref{eq:Bragg-cond-pm-alpha}), which are evidently not a function of $\beta$. These conditions are a function of $\alpha$ and are, therefore, tracked via the same dashed lines in (c). Whenever the two solutions in Eq.~(\ref{eq:Bragg-cond-pm-alpha}) match, these dashed lines intersect resulting in identical lower and upper bandgap frequencies, i.e., a closing of the corresponding bandgap.}
     \label{fig:snapshots_beta}
\end{figure*}

It is also of interest to understand how bandgap limits behave as the value of $\beta$ changes at a given $\alpha$, as shown in Figures~\ref{fig:snapshots_beta}(a) and (b). Bragg bandgaps initiate with a non-zero impedance contrast $\beta$ at frequencies at which the linear dispersion relation folds within the irreducible Brillouin zone, as shown in Figure~\ref{fig:snapshots_beta}(a), and grow in width ($\Delta \Omega$) with higher contrast values. As the contrast $\beta$ approaches the limiting value of unity, the dispersion branches become flat. The growth of $\Delta \Omega$ with increasing magnitude of $\beta$ is further emphasized in Figure~\ref{fig:snapshots_beta}(b), and is shown to be symmetric about $\beta = 0$. It can be seen that even or odd-numbered bandgaps close when the two solutions of Eq.~(\ref{eq:Bragg-cond-pm-alpha}) match regardless of the value of $\beta$. These closings are denoted with dashed lines in Figure~\ref{fig:snapshots_beta}(b).

The behavior is quite different when observing the evolution of bandgap limits with a varying $\alpha$ at specific values of $\beta$, which is depicted in Figure~\ref{fig:snapshots_beta}(c). As the value of $\alpha$ changes, the locus of the bandgap limits oscillates in a manner which increases at higher frequencies. Furthermore, the amplitude (i.e., frequency width) of these oscillations grows as the value of $\beta$ increases.~These oscillatory profiles have nodal points at the locations where the bandgap limit curves intersect, which represent rational values of $\alpha$. At such nodes, the curves corresponding to the Bragg condition established in Eq.~(\ref{eq:Bragg-cond-pm-alpha}), and shown as dotted black lines, also intersect, thus constituting the requisite condition for bandgap closing.

\section{Generalizing bandgap closing conditions to multi-layered PnCs}

\begin{table}[]
\caption{Material properties used in the multi-layered PnC rod unit cells used in Figure~\ref{fig:BG-closing-multi-material-PnC}.}
\centering
\begin{tabular}{l l l }
\hline
Material & Density & Young's Modulus\\
\hline
ABS & 1040 kg/$\text{m}^3$ & 2.4 \ GPa \\
Aluminum & 2700 kg/$\text{m}^3$  & 69 \ \ GPa \\
Brass & 8530 kg/$\text{m}^3$  & 110 GPa \\
Magnesium Alloy & 1800 kg/$\text{m}^3$ & 42 \ \ GPa \\
Steel & 7850 kg/$\text{m}^3$ & 210 GPa \\

\hline
\end{tabular}
\label{table:material_properties}
\end{table}

\begin{figure*}[ht]
     \centering
\includegraphics[]{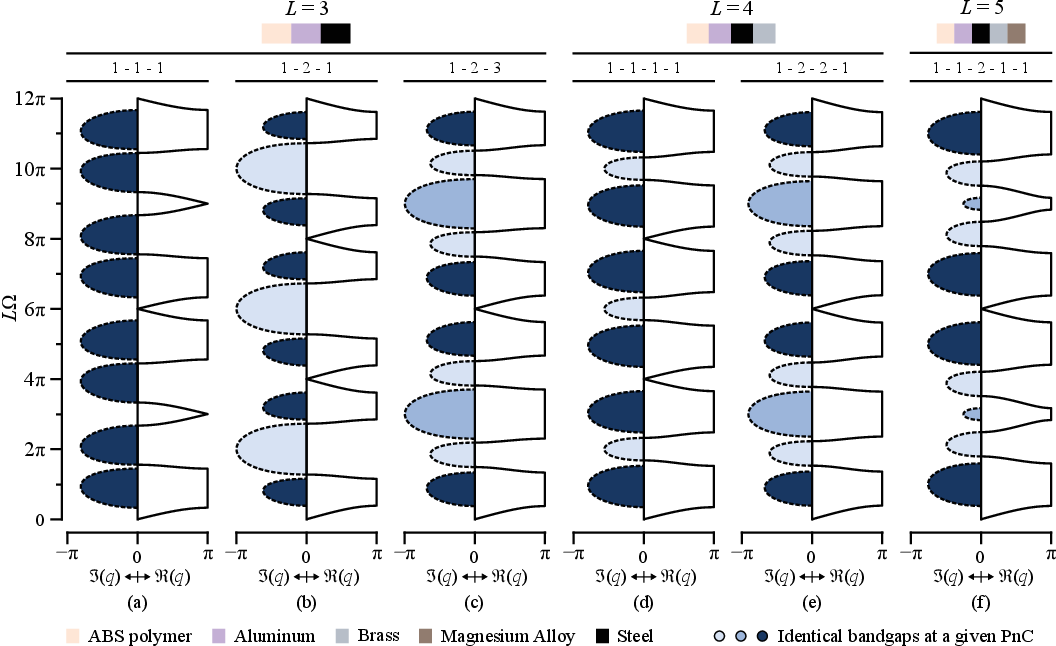}
     \caption{Dispersion diagrams of PnC rods with multi-layered unit cells showing bandgap closings at $L\Omega_p = \pi p\alpha_d$.~(a-c) Three-layered unit cells ($L=3$) with the following $n_1$-$n_2$-$n_3$ harmonic combinations: (a) 1-1-1 (Bandgap closings at $L\Omega_p = 3\pi p$), (b) 1-2-1 (Bandgap closings at $L\Omega_p = 4\pi p$), and (c) 1-2-3 (Bandgap closings at $L\Omega_p = 6\pi p$).~(d-e) Four-layered unit cells with the following $n_1$-$n_2$-$n_3$-$n_4$ harmonic combinations: (d) 1-1-1-1 (Bandgap closings at $L\Omega_p = 4\pi p$) and (e) 1-2-2-1 (Bandgaps closings at $L\Omega_p = 6\pi p$). (f) Five-layered unit cell with the following $n_1$-$n_2$-$n_3$-$n_4$-$n_5$ harmonic combination: 1-1-2-1-1 (Bandgap closings at $L\Omega_p = 6\pi p$). Note that $p\in \mathbb{N}^+$ represents all non-zero natural numbers.}
     \label{fig:BG-closing-multi-material-PnC}
\end{figure*}

While the bandgap closing conditions derived thus far have been mathematically proven for a bi-layered PnC, it is imperative to likewise demonstrate that similar features emerge in a PnC with an arbitrary $L>2$ number of layers.~To generalize our findings, consider a multi-layered unit cell of a PnC rod where all of the constitutive layers have distinct mechanical and geometrical properties. Analogous to the theoretical framework developed earlier, the frequencies corresponding to the Bragg conditions of each of the individual layers can be matched as follows:
\begin{equation}
    n_1 \omega_1 = n_2 \omega_2 = \dots = n_L \omega_L
    \label{eq:equal_nat_freqs}
\end{equation}
which can be alternatively written as:
\begin{equation}
    \frac{\omega_s}{\omega_j} = \frac{n_j}{n_s}
    \label{eq:ratio_freq_Bragg_cond}
\end{equation}
such that $j \neq s$. To locate the frequencies where bandgaps close, we need to pursue a non-dimensional parameter which is reminiscent of the frequency contrast $\alpha$ in the bi-layered PnC. To do so, a generalized harmonic mean can be introduced as follows:
\begin{equation}
    \omega_0 = \left( \frac{\sum_{s=1}^L \frac{1}{\omega_s}}{L} \right)^{-1}
    \label{eq:harmonic_mean_n}
\end{equation}
which can be rearranged to read:
\begin{equation}
\frac{L}{\omega_0} =
    \sum_{s=1}^L \frac{1}{\omega_s} 
    \label{eq:w0_rearranged}
\end{equation}
Next, the frequencies $\omega_s$ can be rewritten as a function of the $j^{\text{th}}$ frequency $\omega_j$ by using Eq.~(\ref{eq:ratio_freq_Bragg_cond}), which, after a few mathematical manipulations, becomes:
\begin{equation}
\frac{1}{\omega_j} = 
\frac{L}{ \omega_0}\frac{n_j}{\sum_{s=1}^L n_s}
\label{eq:freq_layers}
\end{equation}
Note that Eq.~(\ref{eq:w0_rearranged}) is recovered if all the components $1/\omega_j$ in Eq.~(\ref{eq:freq_layers}) are added. Also, it can be clearly seen that the last term in Eq.~(\ref{eq:freq_layers}) is a rational number and, as a result, can be written in a form similar to $\alpha$ as follows:
\begin{equation}
    \frac{\alpha_j}{\alpha_d} = \frac{n_j}{\sum_{s=1}^L n_s}
\label{eq:alpha_multi_layer}
\end{equation}
giving the denominator $\alpha_d=\sum_{s=1}^L n_s$ the same role it played in the bi-layered PnC case, defining the frequency at which a bandgap closes using the equation:
\begin{equation}
    \Omega_p = \frac{1}{L} \pi p\alpha_d
\label{eq:om_p_multi_material_PnC}
\end{equation}
The numerator $\alpha_j = n_j$, on the other hand, provides the number of branches a linear dispersion relation of an individual layer has between Bragg conditions. A couple of additional remarks can be made:

\begin{enumerate}
    \item If $\alpha_d$ is odd, odd and even-numbered bandgaps located at the frequencies given by Eq.~(\ref{eq:om_p_multi_material_PnC}) will close. However, an even $\alpha_d$ will only close even-numbered bandgaps according to the same equation.
    
    \item If the chosen values of $n_s$ have a common factor, a cancellation of the common factor is required for Eq.~(\ref{eq:om_p_multi_material_PnC}) to correctly predict the frequencies at which bandgaps close. For example, if $n_1 = 2$, $n_2 = 4$, and $n_3 = 6$ in a three-layered PnC (i.e., $L = 3$), the number 2 is a common factor. As such, bandgap closing frequencies should be computed using $n_1 = 1$, $n_2 = 2$, and $n_3 = 3$ instead.
\end{enumerate}

Figure~\ref{fig:BG-closing-multi-material-PnC} shows examples of multi-layered PnC with $L = 3, 4, \text{and} \ 5$. The materials used in the three-layered PnC rod are ABS, Aluminum, and Steel in that particular order.~The four-layered and five-layered PnC rods add Brass and Magnesium alloy, respectively.~All material properties are listed in Table~\ref{table:material_properties}. In all cases, the area and total length of all layers are constant and equal to $A_s = 400 \ \text{mm}^2$ and $\ell = 100$ mm, respectively. The individual lengths of the layers are calculated by combining $\ell = \sum_{s=1}^L \ell_s$ and $\omega_s = \ell_s/c_s$ with Eq.~(\ref{eq:equal_nat_freqs}). These equations can be cast into a matrix form as follows:
\begin{equation}
    \begin{bmatrix}
    1 & 1 & 1 & \dots & 1 \\
    \frac{1}{c_1 n_1} & -\frac{1}{c_2 n_2} & 0 & \dots & 0 \\
    \frac{1}{c_1 n_1} & 0 & -\frac{1}{c_3 n_3} & \ddots & \vdots \\
    \frac{1}{c_1 n_1} & \vdots & \ddots & \ddots & 0 \\
    \frac{1}{c_1 n_1} & 0 & \dots & 0 & -\frac{1}{c_L n_L} \\
    \end{bmatrix} 
    \begin{Bmatrix}
    \ell_1 \\ \ell_2 \\ \vdots \\ \ell_L
    \end{Bmatrix}
    =
    \begin{Bmatrix}
    \ell \\ 0 \\ \vdots \\0
    \end{Bmatrix}
\end{equation}

Starting with $L = 3$, we show combinations of the harmonics $n_1$-$n_2$-$n_3$ of (a) 1-1-1, (b) 1-2-1, and (c) 1-2-3. For these PnC configurations, we predict bandgap closings to occur at $\alpha_d=\sum_{s=1}^L n_s$ which correspond to $L\Omega_p = 3\pi p, 4\pi p$, and $6\pi p$ for (a), (b), and (c), respectively, as seen in the respective figures. Similar behavior can be seen in the $L=4$ (1-1-1-1 and 1-2-2-1) and $L=5$ (1-1-2-1-1) cases in (d) through (f). The bandgap width in all cases (a)-(f) has a periodic pattern which repeats itself in between bandgap closings. Additionally, the dispersion branches remain mirror symmetric about the bandgap closings, which is synonymous with the bi-layered ($L=2$) case.

Finally, for completeness, we point out that the existence of two identical layers in a multi-layered unit cell of a PnC rod affects the calculation of $\alpha_j/\alpha_d$ in Eq.~(\ref{eq:alpha_multi_layer}). This is because of a hidden common factor that exists between the chosen vibrational modes that are intended to be matched.~When faced with such a special case, the sum of the modes of the identical layers can be calculated and they can then be treated as a single continuous layer, even if they are not adjacent to one another. For instance, consider a four-layered unit cell made of Aluminum-Steel-Aluminum-Brass.~We choose $n_1 = 1$, $n_2 = 8$, $n_3 = 3$, and $n_4 = 4$ and then calculate $n' = n_1+n_3 = 4$ for a collective mode for the Aluminum layers. As such, the common factor between $n_2$, $n_4$ and $n'$ is 4, and, as a result, $\alpha_d = \sum_s n_s/4$.

\section{Concluding Remarks}

The qualitative and quantitative criteria governing bandgap formation, distribution, and closing conditions were established in a generalized class of rod-based phononic crystals (PnCs) undergoing longitudinal deformations. A transfer-matrix-based approach was used to generate the wave dispersion profiles and develop expressions for bandgap limits and frequencies of maximum attenuation. By implementing two non-dimensional contrast parameters, a frequency contrast $\alpha$ and an impedance contrast $\beta$, which stem from the parameters of the PnC's constitutive layers, the conditions that lead to diminishing bandgaps were derived in closed form, showing that $\alpha$ being a rational number is a necessary condition for bandgap closing in bi-layered PnCs. Furthermore, it was shown that, depending on the parity of the integer numerator and denominator values of $\alpha$, the pattern and frequency location of the bandgap closing can be predicted as a function of the rational number $\alpha$. It was found that the bandgap widths $\Delta \Omega$ of a PnC with a rational $\alpha$ exhibit a periodic profile, which perfectly repeats itself every time a bandgap closes. This pattern was correlated to the resonances of the individual layers of the PnC, and it was proven that matching the natural frequencies of the individual layers (if they were to be treated as stand-alone entities) forces a bandgap to close at the same frequencies. An additional connection was made between bandgap closing criteria and the physics underlying the mode shapes of the individual layers forming the PnC unit cell at different boundary conditions. 

The conclusions drawn from the bi-layered case were generalized to a PnC comprised of three or more layers, where it was similarly shown that the dispersion branches of the multi-layered PnC exhibit mirror symmetry about the frequencies at which bandgaps close. In fully analytical terms, it was also proven that the resonance matching condition for bandgap closing is independent of the number of layers forming the PnC’s unit cell. Resolving the patterns and formation mechanisms of multi-bandgap dispersion profiles is particularly useful for a wide array of new and exciting topics, given the rising interest in exciting applications that require an understanding of how bandgaps close (e.g., topological transition). In tandem, the need to finely tailor phononic band structures remains highly critical for a broad range of elastoacoustic metamaterials. As such, the developments established herein provide a great asset for bandgap engineering in future configurable and tunable PnCs.

\section*{Acknowledgements}

The authors acknowledge the support of this work by the US National Science Foundation through CMMI research award no. 1847254 (CAREER). 

\HA{
\section*{Appendix A: Deriving the dispersion relation from the transfer matrix}
\setcounter{equation}{0}
\renewcommand{\theequation}{A.\arabic{equation}}

When operating under linear reciprocal conditions, the determinant of the transfer matrix is unity \cite{herrero2019matrix}. By combining $|\mathbf{T}-\lambda \mathbf{I}| = 0$, where $\lambda$ is an eigenvalue of $\mathbf{T}$, with $|\mathbf{T}| = 1$, the following characteristic equation is derived:
\begin{equation}
    \lambda^2 - \text{tr}(\mathbf{T}) \lambda + 1 = 0
    \label{eq:ch_eq_TM_UC}
\end{equation}
where, once again, $\text{tr}(\mathbf{T})$ is the trace of $\mathbf{T}$. As a result, the eigenvalues of $\mathbf{T}$ are derived from the roots of Eq.~(\ref{eq:ch_eq_TM_UC}), leading to:
\begin{equation}
    \lambda_\pm = \frac{\text{tr}(\mathbf{T})}{2} \pm \sqrt{\left(\frac{\text{tr}(\mathbf{T})}{2} \right)^2 - 1}
    \label{eq:lambda_roots}
\end{equation}
Examining Eq.~(\ref{eq:lambda_roots}) reveals that $\lambda_- + \lambda_+ = \text{tr}(\mathbf{T})$. Upon experssing the eigenvalues as an exponential function of the wavenumber, i.e., $\lambda_\pm = e^{\pm \mathbf{i} q}$, we arrive at the following dispersion relation:
\begin{equation}
    \text{tr}(\mathbf{T}) = 2\cos(q)
    \label{eq:trace_cos(q)}
\end{equation}
}
\section*{Appendix B: Special case of $\alpha = 0$}
\renewcommand{\theequation}{A.\arabic{equation}}

A special configuration of the bi-layered PnC rod takes place when the frequency contrast $\alpha$ is equal to zero. This is realized when the ratio of the length of layer 1 to that of layer 2 is equal to the sonic speed ratio between the two layers, i.e., 
\begin{equation}
    \frac{\ell_1}{\ell_2} = \frac{c_1}{c_2}
\end{equation}
Upon substituting $\alpha = 0$ in the dispersion relation of Eq.~(\ref{eq:non-dim_disp_rel}), $\Omega$ can be obtained analytically as follows:
\begin{subequations}
\begin{equation}
    \Omega =  k\pi + \frac{1}{2} \cos^{-1} \left(  \beta^2 + (1-\beta^2) \cos(q) \right) 
\end{equation}
\begin{equation}
    \Omega = (k+1)\pi -  \frac{1}{2} \cos^{-1} \left(  \beta^2 + (1-\beta^2) \cos(q) \right) 
\end{equation}
\end{subequations}
for $k \in \mathbb{N}_0$ where $\mathbb{N}_0$ is the set of all natural numbers including zero. The bandgap lower and upper limits can be respectively found using the following expressions:
\begin{subequations}
\begin{equation}
    \Omega_l = k\pi + \frac{1}{2}\cos^{-1} \left(2\beta^2-1 \right)
\end{equation}
\begin{equation}
    \Omega_u = (k+1)\pi - \frac{1}{2}\cos^{-1} \left(2\beta^2-1 \right)
\end{equation}
\end{subequations}
which confirms that all bandgaps have an identical width that is given by:
\begin{equation}
    \Delta \Omega = \pi - \cos^{-1} \left(2\beta^2-1 \right)
\end{equation}

Finally, as stated in Sec.~\ref{sec:bilayered_disp}, setting $\alpha = 0$ in Eq.~(\ref{eq:Max_Att}) shows that $\Omega =r\pi/2$ is the frequency of maximum attenuation inside the bandgap. The attenuation constant corresponding to this frequency is given by:
\begin{equation}
    q_\text{I} = \Im \left[\cos^{-1} \left( \frac{\beta^2+1}{\beta^2-1} \right) \right]
\end{equation}

\bibliographystyle{model1-num-names}
\bibliography{references.bib}











\end{document}